\documentclass{PoS}
\usepackage{subfigure}
\title{Are scalar mesons visible in $B^{\pm} \rightarrow \pi^+\pi^-\pi^{\pm}$ decays?
\thanks{This work has been supported in part by the Polish Ministry of Science
and Higher Education (grant No N N202 248135) and by the IN2P3-Polish
Laboratories Convention (project No 08-127).}}

\ShortTitle{Are scalar mesons visible in $B^{\pm} \rightarrow \pi^+\pi^-\pi^{\pm}$ decays?}

\author{\speaker{L.~Le\'sniak$^{a}$}\\

$^{ a}$ Division of Theoretical Physics, The Henryk Niewodnicza\'nski
Institute of Nuclear Physics, Polish Academy of Sciences, 31-342 Krak\'ow,
Poland
\\
        E-mail: \email{Leonard.Lesniak@ifj.edu.pl}}

\author{J.-P. Dedonder $^{ b}$, A.~Furman$^{ c}$, R.~Kami\'nski$^{ a}$, 
 B.~Loiseau$^{ b}$\\

$^{ b}$ Laboratoire de Physique Nucl\'eaire et de Hautes \'Energies, Groupe Th\'eorie,
Universit\'e Pierre et Marie Curie et Universit\'e Paris-Diderot, IN2P3\&CNRS,
 4 place Jussieu, 75252 Paris, France
$^{ c}$ ul. Bronowicka 85/26, 30-091 Krak\'ow, Poland \\}

\abstract{Two pion effective mass and helicity angle distributions in the charged $B$-meson decays 
into three charged pions are studied. The weak decay amplitudes are calculated in the framework of the 
QCD factorization approximation.
The strong interactions between the pairs of pions are taken into account using scalar and vector pion
form factors. The scalar form factors are constrained by data on pion-pion, kaon-antikaon and four pion
production incorporated into a multichannel model of the coupled amplitudes. The vector form factor is obtained
 from the Belle Collaboration analysis of the $\tau^-\rightarrow \pi^- \pi^0 \nu_{\tau}$ decays.
The theoretical distributions of the dipion effective masses are compared with the corresponding results of the
 recent Dalitz plot analysis of $B^{\pm} \rightarrow \pi^+ \pi^- \pi^{\pm}$ decays performed by the BABAR Collaboration.
 We find that the
 $S$-wave dipion amplitude, although much smaller than the $P$-wave amplitude, 
 plays an important role even in the $\rho(770)$ mass range. We show that the helicity angle distribution 
is strongly asymmetric in the $\rho$- meson range.
 This effect can be attributed to the broad $f_0(600)$(or $\sigma$) meson.  The fact that the signal of the 
$B^{\pm} \rightarrow f_0(980) \pi^{\pm}$ decay has not been found in the experimental analysis can be easily explained
 in our
 model since the relevant $B$ decay amplitude is proportional to the scalar form factor which has a  
dip at the $f_0(980)$ mass. We obtain a unified unitary description of the contribution of the three scalar 
 resonances $f_0(600)$, $f_0(980)$ and $f_0(1400)$ in terms of the pion non-strange scalar form factor.
}

\FullConference{35th International Conference on High Energy Physics,\\
		 July 22 - 28 2010\\
		 Paris, France}

\begin{document}

\section{Introduction}
Studies of the charged $B$ mesons decays into three charged pions 
constitute an important part of the analyses of the three-body charmless hadronic
$B$ decays (see, for example~\cite{El-Bennich2009} and references given therein).
 Investigation of the Dalitz diagrams with  theoretically well constrained
 meson-meson strong interaction amplitudes can lead to a better extraction of the weak interaction
parameters from experimental data. Recently the BABAR Collaboration has published results of 
an isobar model analysis of the $B^{\pm}$ decays into $\pi^{\pm}\pi^{\mp}\pi^{\pm}$
~\cite{Aubert:2009}. The $\pi^+\pi^-$ effective mass spectrum is largely dominated by the 
$\rho(770)$ meson, yet the authors of Ref.~\cite{Aubert:2009} find some contribution of the scalar 
resonance $f_0(1370)$ but none of the $f_0(980)$. Here we present the results of a theoretical
model in which the $S$-wave pion-pion strong interaction amplitude is constrained
using a multichannel unitary approach including the $\pi^+\pi^-$ coupling to $K^+K^-$, 
$K^0\bar K^0$ 
and $4\pi$ (effective $(2\pi)(2\pi)$) states. Application of this model to the partial wave 
analysis of the BABAR data shows that the scalar meson contributions, in particular that of the 
$f_0(600)$, are needed to explain the effective $\pi\pi$ mass distributions.

\section{Theoretical model}
The weak decay amplitudes of $B^{\pm}$, corresponding to the quark transitions 
$b \rightarrow u\bar u d$ and $b \rightarrow d\bar d d$, are derived in the QCD quasi-two body
factorization approach for the limited range of the effective $\pi^+ \pi^-$ masses less than 
about 1.7 GeV. One assumes that only two of the three produced pions interact strongly, forming 
either an $S$- or $P$-wave state denoted by $R_S$ or $R_P$. The $\pi\pi$ strong interaction 
amplitudes are constrained by chiral symmetry, by QCD and by experimental data on
meson-meson interactions. The matrix elements of the effective weak Hamiltonian
involve the pion non-strange scalar and vector form factors.
 
The $\pi^-(p_1) \pi^+(p_2)$ $S$-wave contribution to the 
$B^- \to \pi^-(p_1) \pi^+(p_2) \pi^-(p_3)$ decay amplitude reads:
\begin{equation}
\label{MSsij}
\mathcal{M}^-_S(s_{12})=\frac{G_F}{\sqrt{3}}\left[-\chi_S f_\pi \left(M_{B}^2-s_{12}\right)
F_0^{BR_S}(m_\pi^2) u(R_S \pi^-) 
  + B_0 \frac {M_B^2-m_\pi^2}{m_b-m_d} 
F_0^{B\pi}(s_{12}) v(\pi^-R_S)\right] \Gamma_1^{n^*}(s_{12}), 
\end{equation} 
where
\begin{eqnarray}
\label{RSpi}
u(R_ {S} \pi^-)&=&\lambda_u\left[a_1(R_ {S}\pi^-)
  +  a_4^u(R_{S}\pi^-)  +a_{10}^u(R_{S}\pi^-)-\left( a_6^u(R_{S}\pi^-) +a_8^u(R_{S}\pi^-)\right)
  r_\chi^\pi\right] \nonumber\\
 &&+ \lambda_c\left[
    a_4^u(R_{S}\pi^-)  +a_{10}^c(R_{S}\pi^-)-\left( a_6^c(R_{S}\pi^-) +a_8^c(R_{S}\pi^-)\right)
  r_\chi^\pi\right], 
\end{eqnarray}
\begin{equation}
\label{piRS}
v(\pi^- R_S)=\lambda_u  \left [-2a_6^u(\pi^-R_S)+a_8^u(\pi^-R_S)\right]
+ \lambda_c \left [-2a_6^c(\pi^-R_S)+a_8^c(\pi^-R_S)\right] 
\end{equation}
and
\begin{equation}
 B_0=\frac{m_{\pi}^2}{m_u+m_d},~~~~~~~~~~~~~~~~~~r_\chi^\pi=\frac{2B_0}{m_b+m_u}.
\end{equation}

In the above equations $G_F$ denotes the Fermi coupling constant, 
$f_{\pi}$ is the pion decay constant,
  $s_{12}=(p_1+p_2)^2$, $p_1$ and $p_2$ being the $\pi^-$ and $\pi^+$ momenta.
The masses of the charged $B$ mesons, of the charged pions, of the $b,~ u$ and $d$ quarks are denoted by 
$M_B$, $m_\pi$, $m_b,\ m_u$ and $\ m_d$, respectively. The fitted parameter $\chi_S$ is the
 proportionality 
factor which appears under the assumption that the decay amplitude of $R_S$ state into two pions
is proportional to the pion non-strange scalar form factor $\Gamma_1^{n^*}(s_{12})$.
The $BR_S$ and $B\pi$ transition form factors are denoted by $F_0^{BR_S}(m_\pi^2)$ and 
$F_0^{B\pi}(s_{12})$, respectively. The symbols 
$\lambda_u= V_{ub} V^*_{ud}$ and $\lambda_c= V_{cb} V^*_{cd}$ are products of the 
Cabibbo-Kobayashi-Maskawa quark-mixing matrix elements $V_{qq'}$. The effective
 Wilson coefficients $a^{u,c}_j$, $j=1,4,6,8,10$,
are calculated to next-to-leading order in the strong coupling constant
including the vertex and penguin corrections. 

The pion non-strange scalar form factor is calculated in a unitary relativistic three 
coupled-channel model using  the $\pi \pi$, $K \bar K$ and effective $(2\pi)(2\pi)$ scattering
 $T$ matrix of Refs.~\cite{Kaminski:1997gc}. It is constrained at low energy by chiral 
perturbation theory.
This form factor depends on two fitted parameters: the first one insures the 
convergence of the involved integrals and the second one controls the high-energy 
behaviour of $\Gamma_1^{n^*}(s_{!2})$.
 
Explicit expressions for the $P$-wave contributions to the $B$ decay amplitudes can be found in
~\cite{ph2010}. Here the pion vector form factor takes into account the contributions of the
three vector resonances $\rho(770)$, $\rho(1450)$ and $\rho(1700)$ and follows from the Belle 
Collaboration analysis of the semi-leptonic $\tau^- \to \pi^- \pi^0 \nu_\tau$ decays~\cite{Fujikawa_PRD78_072006}.
For the $P$-wave amplitude we introduce a fitted overall normalization factor $N_P$. 
 Due to presence of two identical pions in the final state
one has to symmetrize the decay amplitudes over the two possible $\pi^+\pi^-$ combinations.

\section{Results and discussion}

We obtain a good fit to the $\pi \pi$ effective mass distributions of the BABAR Collaboration 
data of the $B^\pm \to \pi^\pm \pi^\mp \pi^\pm$  decays~\cite{Aubert:2009}.
The value of the theoretical branching fraction for the $B^\pm \to \rho(770)^0 \pi^\pm$ decays,
 $(8.2 \pm 0.5\times) 10^{-6}$, agrees well with that,
 $(8.1\pm0.7\pm1.2^{+0.4}_{-1.1})\times 10^{-6}$, of the experimental analysis.
The normalization factor $N_P$ is found to be close to 1.

In Fig.~1 we show the $\pi^+\pi^-$ effective mass distributions for the $B^+$ decays.
 In the left plot 
the value of the cosinus of the pion helicity angle $\theta$ is negative and in the right one it is 
positive.
The $\pi^+ \pi^-$ spectra are dominated by the $\rho(770)^0$ resonance but at low effective
masses the $S$-wave contribution is sizable.
Here the $f_0(600)$ resonance manifests its presence.
Furthermore one observes a strong negative or positive interference of the $S$ and $P$ waves in 
the event 
distributions depending on the sign of  $cos \theta$. The interference term between the $S$ and $P$ dipion 
amplitudes can reach a value as high as 30\%
of the dominating $\rho(770)$ contribution.
The $f_0(980)$ resonance is not directly visible as a peak, since the scalar form factor has 
a dip near 1~GeV.
At 1.4~GeV the maximum of the $S$-wave distribution comes from the scalar resonance 
$f_0(1400)$~\cite{Kaminski:1997gc}. 
Integrating the overall $S$-wave contribution in the whole
 available phase space one obtains as much as 25 \% of the total 
$B^\pm \to \pi^\pm \pi^\mp \pi^\pm$ branching fraction.
Similar results to those shown in Fig.~1 are obtained 
for the $B^-$ decays. The absolute differences between the $B^+$ and $B^-$ distributions are small so the 
corresponding $CP$ asymmetry is also small, about 3\%.

Our model yields a unified description of the contribution of the three scalar resonances
 $f_0(600)$, $f_0(980)$ and $f_0(1400)$ in terms of one function: the pion non-strange scalar 
form factor.
This reduces strongly the number of needed free parameters to analyze the Dalitz plot.
The functional form of our $S$-wave amplitude, proportional to
$\Gamma _1^{n^*}(s)$, could be used in Dalitz-plot analyses and the table of
$\Gamma _1^{n^*}(s)$ values can be sent upon request.

Let us note that the strong interaction phases of the decay amplitudes are constrained by
 unitarity and meson-meson data. This should help to improve the
 precision of the weak interaction amplitudes extracted from Dalitz plot analyses, especially
the value of the weak angle phase $\gamma$ (or $\phi_3$).
Of course new experimental data with better statistics would be welcome.
One expects new results on the $B^\pm \to \pi^\pm \pi^\mp \pi^\pm$ decays from the Belle 
Collaboration and in near future, from LHCb and super $B$ factories.

In summary, we have shown that the scalar mesons are clearly visible in the
 $B^{\pm} \rightarrow \pi^+\pi^-\pi^{\pm}$ decays. Moreover, the introduction to the decay amplitudes of the pion non-strange scalar form 
factor, constrained by theory and other experiments than $B$ decays, can be more economical in the description
of data than the use of the isobar model with many free parameters fitted for each scalar resonance and,
in addition, for the nonresonant term. 


\begin{figure}[!ht]
\subfigure{\includegraphics[width=0.45\textwidth]{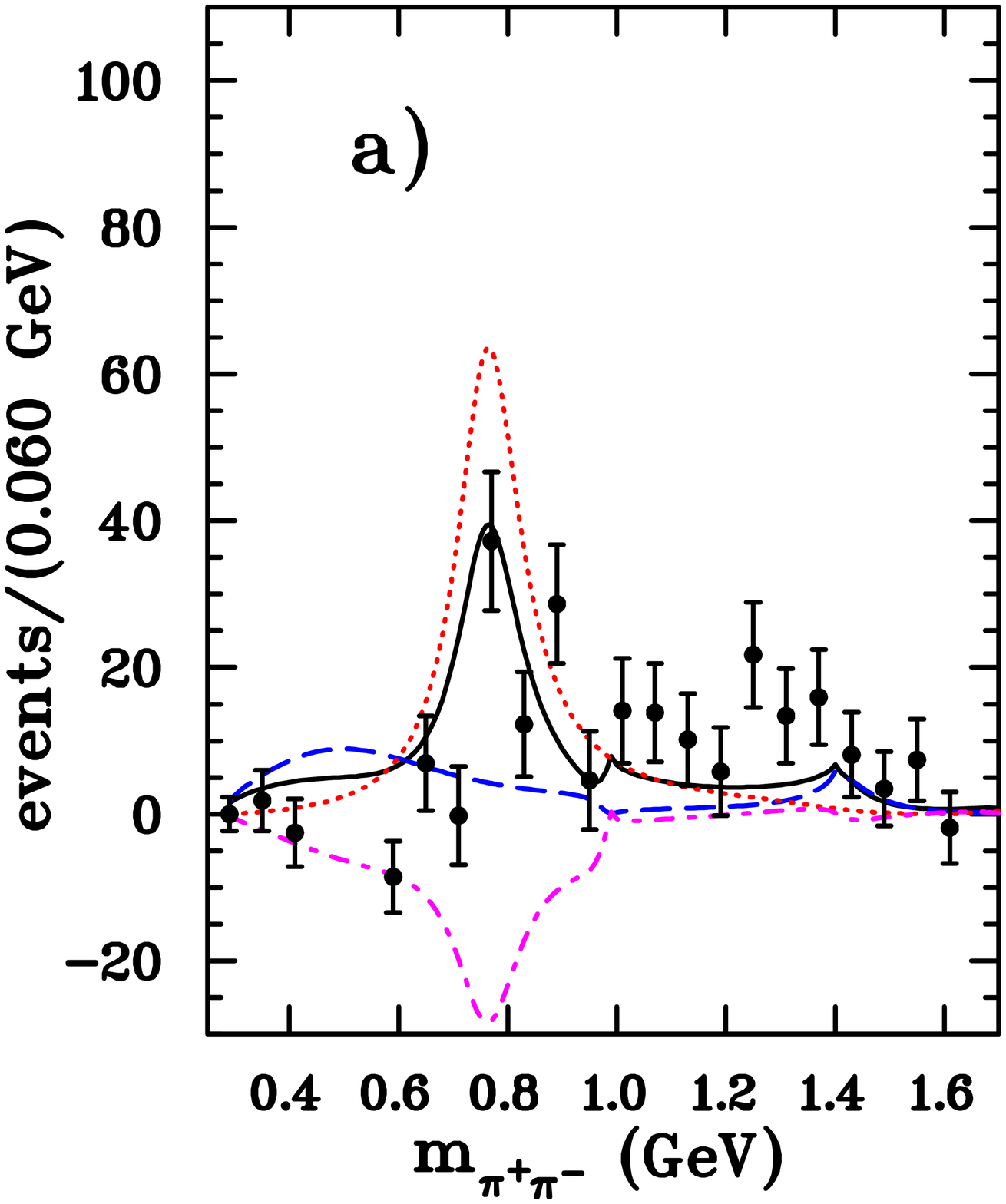}}~~~~~
\subfigure{\includegraphics[width=0.45\textwidth]{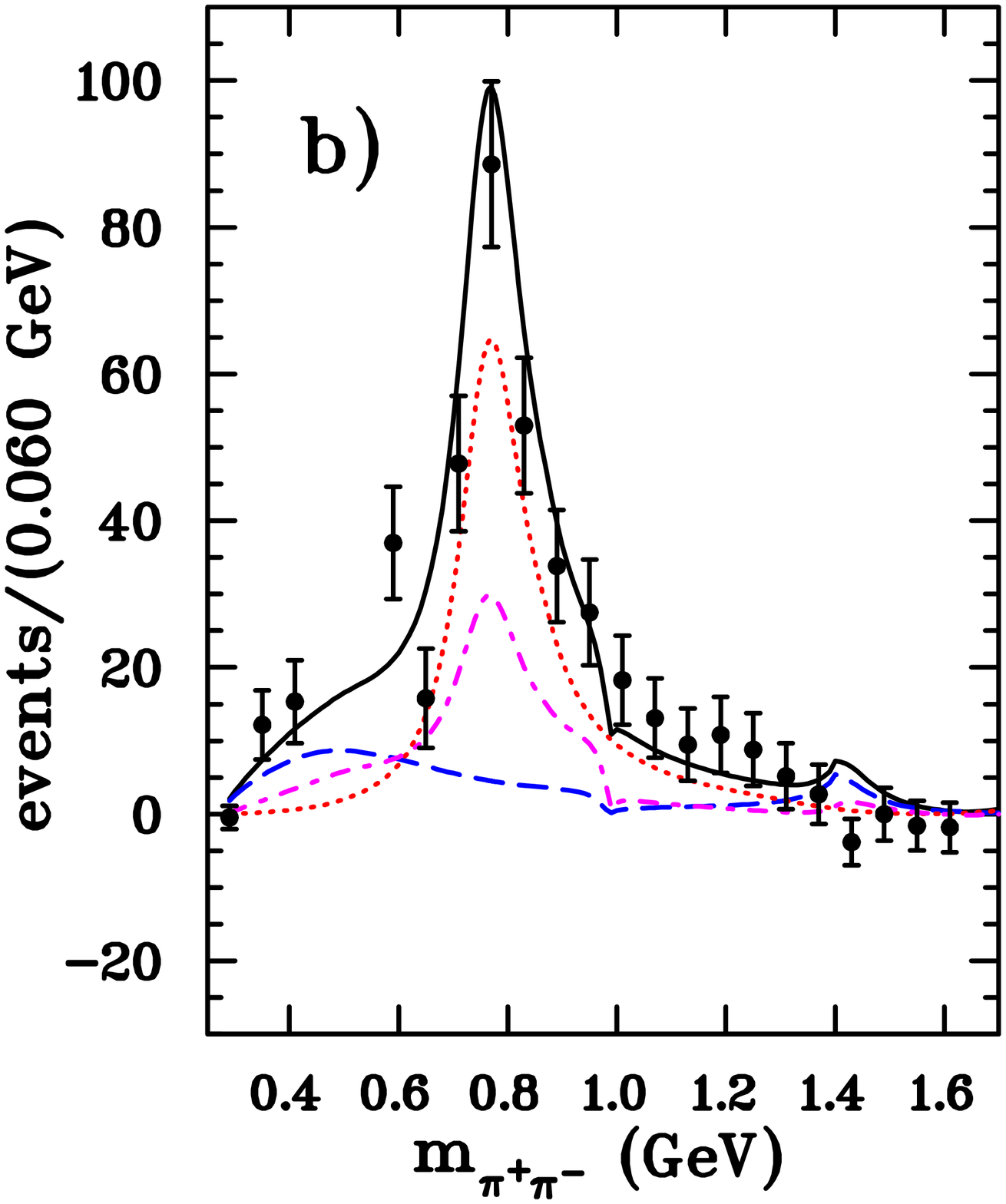}}
\caption {The $\pi^+\pi^-$ light effective mass distributions from the fit to the BABAR experimental
 data~\cite{Aubert:2009} for the $B^+$ decays into $\pi^+ \pi^- \pi^+$  a) with $cos \theta <0$ and  b) with 
$cos \theta >0$, where $\theta$ is the helicity angle. 
The dashed line represents the $S$-wave contribution, the dotted line 
that of the $P$ wave and the dot-dashed line that of the interference term. 
The solid line corresponds to the sum of these contributions.}

\end{figure}




\end{document}